\documentclass[showpacs, prb, twocolumn]{revtex4}
\usepackage{latexsym}
\usepackage{amsfonts}
\usepackage{amssymb}
\usepackage{amsmath}
\usepackage{graphicx}

\begin{document}

\preprint{APS/123-QED}

\title{Spin accumulation probed in multiterminal lateral all-metallic devices}
\author{M. V. Costache}
\email{m.v.costache@rug.nl}

\affiliation{Department of Applied Physics and Materials Science
Center, University of Groningen,\\
Nijenborgh 4, 9747 AG Groningen, The Netherlands}

\author{M. Zaffalon}
\altaffiliation[present address: ]{Dept. of Condensed Matter,
Weizmann Institute of Science, 76100 Rehovot, Israel}

\author{B. J. van Wees}
\affiliation{Department of Applied Physics and Materials Science
  Center, University of Groningen,\\
Nijenborgh 4, 9747 AG Groningen, The Netherlands}

\begin{abstract}
We study spin accumulation in an aluminium island, in which the
injection of a spin current and the detection of the spin
accumulation are done by means of four cobalt electrodes that
connect to the island through transparent tunnel barriers.
Although the four electrodes are designed as two electrode pairs
of the same shape, they nonetheless all exhibit distinct
switching fields. As a result the device can have several
different magnetic configurations. From the measurements of the
amplitude of the spin accumulation, we can identify these
configurations, and using the diffusion equation for the spin
imbalance, we extract the spin relaxation length
$\lambda_\mathrm{sf} = 400 \pm 50$~nm and an interface spin
current polarization $P = (10 \pm 1)\%$ at low temperature and
$\lambda_\mathrm{sf} = 350 \pm 50$~nm, $P = (8 \pm 1)\%$ at room
temperature.
\end{abstract}


\maketitle

It is an interesting question what happens to the transport
properties in non-magnetic conductors if the carriers are spin
polarized. This is both a fundamental question in the field of
spintronics and has also practical applications \cite{rew}. In
an all-electrical setup, spin polarized carriers are injected by
driving a current from a ferromagnet. This induces an imbalance
between the two spin populations, that, for diffusive systems,
extends over a distance of order $\lambda_\mathrm{sf} =
\sqrt{D\tau_\mathrm{sf}}$ from the interface. $\tau_\mathrm{sf}$
is the spin lifetime and $D$ the electron diffusion constant for
the conductor. If a second ferromagnet is present within
$\lambda_\mathrm{sf}$ from the injector, it can be used to
detect the spin accumulation.

In order to study spin related transport in a non-magnetic metal
using a lateral device, a true multi-terminal device is needed.
By spatially separating the current path from the voltage
probes, one can distinguish between truly spin-related effects
and spurious, interface-dependent phenomena.

This technique, pioneered by \citet{johnson_85}, has been
successfully extended to the study of spin transport in diverse
systems, from metallic systems at low and room temperature
\citep{jedema_01, jedema_02, urech_04, Otani, Zutic02} to carbon
nanotubes \citep{niko}, and to a lesser extent, in
semiconductors \citep{ohno}, superconductors \citep{beckmann}
and organic materials \citep{dediu}. In the case of metallic
systems, the interface between the ferromagnet and the metal has
been varied from transparent to tunnelling. \citet{valenzuela01}
have used a lateral spin valve device to probe the magnitude and
sign of the polarization of a ferromagnetic contact as a
function of the injecting bias voltage.

\begin{figure}
\includegraphics[width=6cm]{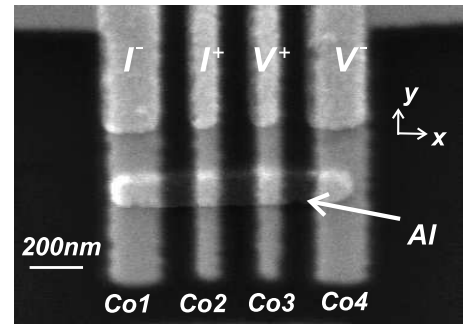}
\caption{Scanning electron microscope (SEM) image of the device.
Visible are the Al island and the four Co contacts of two
different widths: the wider electrodes connecting the island at
its ends, have a lower switching field. In the non-local
measuring configuration, a current $I$ is injected from $Co2$ to
$Co3$ and the voltage difference $V = V^{+} - V^{-} $ is
detected between $Co3$ and $Co4$. All measurements presented in
this article are taken in the non-local configuration.}
\label{sample}
\end{figure}

Recently, the spin accumulation in a diffusive Al island, with
all lateral dimensions smaller than $\lambda_\mathrm{sf}$ has
been studied \citep{zaffalon_03}. The island is contacted by
four Co electrodes via tunnel barriers for injection/detection
of the spin accumulation. However, this system suffers from
several drawbacks such as difficulty of fabrication and, more
importantly, large magnetic fringing fields at the end of the
electrodes, which can affect the spin accumulation. Also it is
not straightforward to reduce the island's volume to increase
the spin accumulation.

In this work, we focus on a 1-D system (only one lateral
dimension larger than $\lambda_\mathrm{sf}$) in which an Al
island is contacted with four in-line Co electrodes, as shown in
Fig.~\ref{sample}.

The orientation of the electrodes' magnetization is pinned along
the electrode axis in the substrate plane by the shape
anisotropy and can be switched by an external magnetic field in
the $\hat{\mathbf{y}}$-direction. The inner/outer electrode
pairs are designed to have different widths.

As the switching field is lower for the wider (outer)
electrodes, we have a control on the magnetization of the
individual electrodes. However, we will see that the switching
fields for identically designed Co electrodes may not be the
same. This is due to the small differences produced during the
fabrication and to magnetic interactions between the electrodes
ends.

Here, we study the spin accumulation as a result of the
different orientations of the four Co electrodes and we show
how, from the magnitude of the spin accumulation, we can infer
the magnetic configuration of the electrodes, as well as the
polarization of the Co/Al$_2$O$_3$/Al contacts and the spin
diffusion length $\lambda_\mathrm{sf}$ in Al.

The theoretical analysis of the spin imbalance in our Al strip
is based on the model for diffusive transport introduced by
\citet{van_son_87, johnson_88} and refined by \citet{valet_93}:
there transport was analyzed for transparent
ferromagnetic/non-magnetic (FM/N) interfaces. It was later
understood \citep{filip} that the efficiency of the injection,
i.e. the ratio spin polarized current to total current, can be
increased by interposing a spin dependent interface between FM
and N, such as a high resistance tunnel barrier.

The devices (see Fig.~\ref{sample}) are made by electron beam
lithography and two-angle shadow mask evaporation process. The
shadow mask consists of a PMMA-MA/germanium/PMMA try-layer, the
base resist having higher sensitivities than the top resist as
to enable, by selective exposure, the making of a suspended mask
with large undercuts. First, we deposit through the suspended
mask 20~nm thick Al at 35$^\circ$ on the Si/SiO$_2$ substrate
using electron-gun evaporation to form 1~$\mu$m$\times 150$~nm
strip. Next, we expose Al to pure oxygen at a pressure of
$10^{-2 }$~mbar for few minutes to form a thin Al$_{2}$O$_{3}$
layer. In the last step, four Co electrodes 30~nm thick are
deposited perpendicular to contact the Al strip, without
breaking the vacuum. The resistance of the Al/Al$_{2}$O$_{3}$/Co
tunnel junctions ranges from $20-60~\Omega \cdot \mu$m$^2$,
depending on the oxidation time.

As mentioned above, the inner/outer Co electrodes have been
designed to have different widths, with the outer contacts at
150~nm and the inner ones at 80~nm. This allows us to
independently flip the magnetization direction of the
electrodes, when an external magnetic field is slowly swept
($\approx$ 1--2~mT/sec), along the contacts' direction.
Nevertheless, we will present measurements in which sometimes
the narrow contacts switch at lower fields than the wider ones.

Measurements were performed at about 2~K and at room temperature
by standard a.c. lock-in techniques, with a modulation frequency
of 7--17~Hz. We have measured 5 devices in detail.

\begin{figure}
\includegraphics[width=8cm]{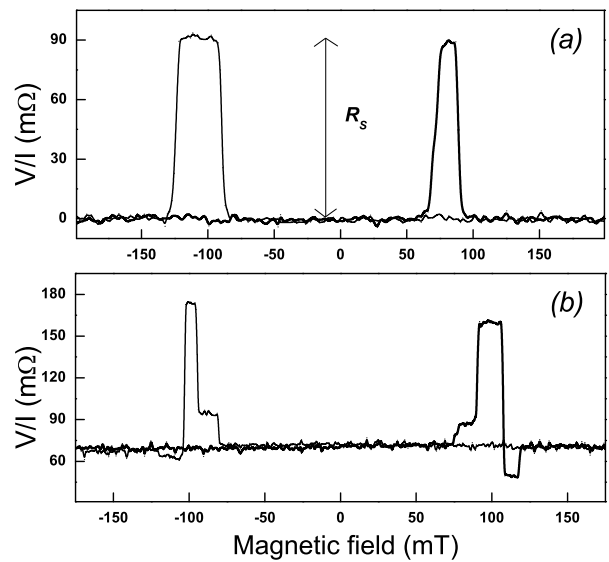}
\caption{Non-local spin valve measurement: the transresistance
$V/I$ as a function of the in-plane magnetic field for positive
and negative sweep direction. (a) Device A, two switch traces of
the cobalt electrodes and (b) Device B, four switch traces, at
low temperature (2~K).} \label{complete}
\end{figure}

All measurements presented here are taken in the \emph{non
local} measuring configuration: a current $I$ is injected from
Co2 to Co1 and a voltage $V$ is detected between Co3 and Co4.
Since no charge current flows through the voltage detectors, our
device is not sensitive to interface or bulk magnetoresistance
related effects, but only to the spin degree of freedom.

Figure \ref{complete} shows two typical nonlocal spin valve
measurements for different devices at low temperature. The
plotted signal is $(V^{+} - V^{-}) / I$, as a function of the
in-plane (in the $\hat{\mathbf{y}}$ direction) magnetic field.
Referring to Figure \ref{complete}(a) device A, at +200~mT, all
contacts' magnetization are aligned parallel to the external
magnetic field, in the $+\hat{\mathbf{y}}$ direction. We sweep
the magnetic field toward negative values. At --80~mT, the two
larger electrodes, namely Co1 and Co4, flip their magnetization
(antiparallel configuration), and the detected signal increase
to 90~m$\Omega$ above a zero background. Upon increasing the
magnetic field further to --120mT, the two smaller electrodes,
Co2 and Co3, flip, the magnetization of the four contacts is
parallel again, but now in the opposite direction
($-\hat{\mathbf{y}}$). The reverse trace show a similar
behavior. Also for repeated sweeps, the field at which the
magnetization switching occurs is within 20~mT of the given
values.

Now, what happen if all four Co electrodes switch their
magnetization at different fields? Figure \ref{complete}(b)
shows such a measurement for device B. We interpret the
additional steps in the signal as the fingerprint of different
magnetic configurations of the electrodes. Again, at --170~mT,
we start with a parallel configuration of the Co electrodes and
with a background level of +70~m$\Omega$ (the nature of which is
unknown). Ramping the field to positive values, at +75~mT, Co1
flips, the injectors are antiparallel and the signal increase
above the background level by +17~m$\Omega$. At +90~mT the other
largest electrode, Co4, reverses, so that also the detectors are
antiparallel to each other. The spin signal increases now by
+90~m$\Omega$ above the background level. At 106~mT, Co3 strip
flips, the detectors return to parallel and the signal drops by
20~m$\Omega$ below the background level. The electrodes stay in
this configuration until reaching a field of +120~mT, when the
other smallest electrode, Co2, switches and the signal reaches
the background level. The sweep to negative fields shows a
similar behavior, with a difference in the value of the spin
signal, probably due to presence of the magnetic domains in the
Co strips.

It is worth mentioning at this point that, given the symmetric
positions of the electrodes on the island, one cannot tell
whether the electrode flipping at the lowest field, for example
in Fig.~\ref{complete}(b), is Co1 or Co4. This uncertainty could
have been avoided, for instance if the electrodes were arranged
in a wide, narrow, wide, narrow fashion (and if the switching
field is determined by the lateral dimension of the electrode
only).

\begin{figure}
\includegraphics[width=8cm]{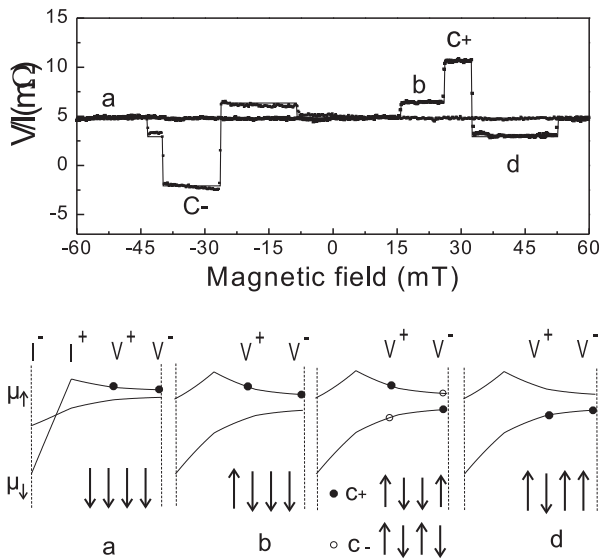}
\caption{(Top) Experimental data (dots) and fitting results
(lines) using eq.1 for nonlocal spin valve at room temperature,
device C. The letters \emph{a} to \emph{d} represent the
different magnetic configurations as described below. (Bottom)
Spatial dependence of $\mu_\uparrow$ and $\mu_\downarrow$
electrochemical potentials in the Al island for the magnetic
configurations \emph{a} to \emph{d}, as in the top panel. The
filled (open) dots indicate the potential measured by the
$V^{+}$ and $V^{-}$ probes.} \label{chem}
\end{figure}

Figure 3 top panel shows data for device C measured at room
temperature. The behavior is similar to that of device B. The
spin signal of 6-7~m$\Omega$ is smaller due to a lower spin
relaxation length, and a somewhat smaller interface polarization
at room temperature. For both positive and negative sweep
directions of the magnetic field, we identify five magnetic
configurations, \emph{a}, \emph{b}, \emph{c+}, \emph{c--} and
\emph{d}.

To clearly illustrate the spin contributions in different
magnetic configurations, we refer to Fig.~\ref{chem} (bottom).
Here we show schematically the spatial dependence of the spin-up
($\mu_\uparrow$) and spin-down ($\mu_\downarrow$) chemical
potentials in the Al island, for the different magnetic
configurations, when a charge current is injected from $I^{+}$
to $I^{-}$. $V^{+}$ and $V^{-}$ represent the position of the
voltage probes. Let us assume, for the moment, the contacts to
be 100\% spin polarized: $V^{+},V^{-}$ would detect either
$\mu_\uparrow$ or $\mu_\downarrow$, according to the
magnetization direction of the contact.

In the configuration \emph{a} in which all contacts are
parallel, and the spin related signal arises from the spatial
dependence of $\mu_\downarrow(x)$.

When Co1 flips, configuration \emph{b}, the injectors are
antiparallel, a non-uniform spin accumulation is present in the
Al island, and relaxes from the points of injection, giving rise
to a spin current $I_s \propto \sigma_N \cdot \nabla
(\mu_\uparrow - \mu_\downarrow)$. Note that the charge current
$I \propto \nabla (\mu_\uparrow + \mu_\downarrow)$ at Co3 and
Co4 is absent. Although the detectors are still parallel and
sensitive only to spin down channel, the signal is somewhat
larger than in configuration (a), as it can be seen in the
measurement, by 1.6~m$\Omega$.

When Co4 reverses, configuration \emph{c+} (black dots), also
the detectors are antiparallel, the $V^{+}$ electrode detects
$\mu_\downarrow$ and $V^{-}$, $\mu_\uparrow$. In this
configuration, with both injectors and detectors antiparallel to
each other, we obtain the highest spin contribution, that is
6~m$\Omega$ in our measurement. When also Co3 flips,
configuration \emph{d}, the detectors now measure the spatial
dependence of $\mu_\uparrow$, so that the magnitude of the
signal is the same as in configuration \emph{b} but with
opposite sign. In the reverse trace, configuration \emph{c--}
(open dots) the notable difference is that Co3 flips before Co4
and the signal changes sign as $V^{+}$ is sensitive to spin up
while $I^{+}$ injects spin down electrons.

To evaluate qualitatively the experimental results, we model the
system as i) one dimensional and we assume injectors and
detectors to be ii) collinear (parallel or antiparallel to
$\hat{\mathbf{y}}$), and iii) point-like. Assumption i) and iii)
are justified by the fact that previous measurements reported
$\lambda_\mathrm{sf} = 500$~nm at RT, \citep{jedema_02,
zaffalon_03}, larger than the island's and contacts' width, and
ii) because shape anisotropy keeps the magnetization in-plane
and in the direction of the contact. The contacts' positions of
Co1, Co2, Co3, Co4 electrodes to the Al island are at $d_1$,
$d_2$, $d_3$ and $d_4$, with $0 \leq d_1 < d_2 < d_3 < d_4 \leq
L$. A charge current $I$ is injected at $d_2$ and extracted at
$d_1$. As the injectors are ferromagnetic, the injected charge
current is partially spin polarized, $I_s = P_i I$, ($P < 1$ and
$i = 1, 2$). This produces a space dependent spin accumulation
in the Al island $\mu(x) = \tilde{\mu}(x, d_2) - \tilde{\mu}(x,
d_1)$ (the minus sign because of the opposite directions of the
charge current) where $\tilde{\mu} = (\mu_\uparrow -
\mu_\downarrow)/2$. The spatial dependence of $\mu(x)$ in Al
strip, can be calculated by solving the 1-D spin coupled
diffusion equation \citep{van_son_87} with the boundary
conditions $\mathrm{d}\mu_{\uparrow, \downarrow} / \mathrm{d}x =
0$ at either ends of the island $x = 0$ or $x = L$, that is, no
charge or spin current at $x = 0, L$. The solution is
\citep{zaffalon_thesis}

\begin{equation}
\begin{split}
\tilde{\mu}(x,d_{i})=\frac{e \lambda_\mathrm{sf}I P_{i}
}{2\sigma A}
\bigg[ \exp \left( -\frac{\vert x-d_{i}\vert}{\lambda_\mathrm{sf}}\right) \\
+ C_{i} \exp \left( -\frac{x}{\lambda_\mathrm{sf}} \right) +
D_{i} \exp \left(\frac{x-L}{\lambda_\mathrm{sf}} \right) \bigg]
\end{split}
\label{bigeq}
\end{equation}
where $\sigma$ and A are the conductivity and sectional area of
the Al strip and $C_{i}$ and $D_{i}$ are given by

\begin{equation}
C_{i}=\frac{\cosh[(L-d_{i})/\lambda_\mathrm{sf}]}{\sinh[L/\lambda_\mathrm{sf}]}\;
\hspace{3mm} D_{i}=
\frac{\cosh[d_{i}/\lambda_\mathrm{sf}]}{\sinh[L/\lambda_\mathrm{sf}]}
\label{const}
\end{equation}

In the limit $L \gg \lambda_\mathrm{sf}$, one recovers the 1-D
equation \citep{jedema_02} and in the limit $L
\ll\lambda_\mathrm{sf}$, one finds the 0-D expression
\citep{zaffalon_03}. The ferromagnetic detectors positioned at
$d_3$ and $d_4$ have polarization $P_3$ and $P_4$. We measure
the difference of the detectors' potentials at this points,
$V(d_3) - V(d_4) = V^{+} - V^{-}$, and the spin dependent
resistance is $(V^{+} - V^{-}) / I = [P_{3}\mu(x = d_{3}) -
P_{4}\mu(x = d_{4})] / eI$.

In the fitting, the free parameters are $|P_i| = P$ and
$\lambda_\mathrm{sf}$, the position of the electrodes are as
determined from the SEM micrographs. Also, a constant background
is added to the calculated signal. We find at low temperature a
spin diffusion length $\lambda_\mathrm{sf} = 400 \pm 50$~nm and
an interface polarization $P = (10 \pm 1)\%$ and
$\lambda_\mathrm{sf} = 350 \pm 50$~nm, $P = (8 \pm 1)\%$ at room
temperature. Both these values are slightly smaller than
previously found \cite{jedema_02, valenzuela01, zaffalon_03}.

\begin{figure}
\includegraphics[width = 0.5\textwidth]{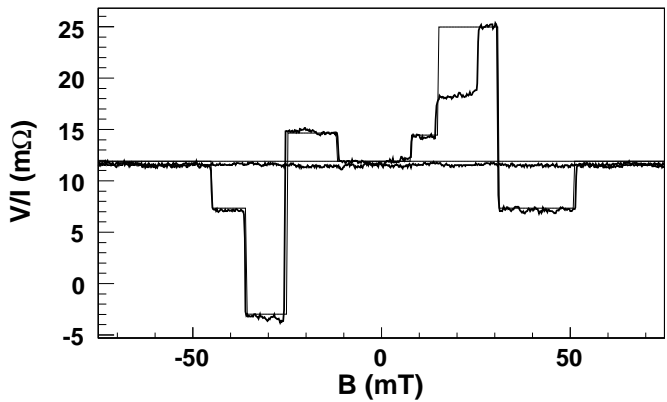}
\caption{Non-local spin accumulation at room temperature, device
D. The fitting based on eq.~(\ref{bigeq}) returns
$\lambda_\mathrm{sf} = (350 \pm 50)$~nm and a polarization of $P
= (8 \pm 1)\%$.} \label{five_switching}
\end{figure}

Figure~\ref{five_switching} shows a non-local spin valve
measurement at room temperature for device D. We observe, while
sweeping at positive fields, at around +14~mT, an extra step in
the signal and we interpret this as Co4 flipping its
magnetization through an intermediate step. We also note in the
reverse trace that Co3 reverses before Co4. Also here, the fit
follows well the experimental data: this implies that all
junctions have the same polarization.

In summary, we have studied spin accumulation in an Al island,
connected by four Co electrodes through low resistance
junctions. From the measurements of the amplitude of the spin
accumulation we can identify the sequence of the magnetization
switching of the ferromagnetic contacts. The analysis based on
eq.~(\ref{bigeq}) allows us to extract $\lambda_\mathrm{sf}$ and
$P$.

This work was supported by the Stichting Fundamenteel Onderzoek
der Materie (FOM) and MSC$^\textrm{plus}$. We acknowledge useful
discussions with Andrei Filip and Steven M. Watts and the
technical support of Gert ten Brink and Pim van den Dool.

\bibliographystyle{apsrev}
\bibliography{ref}

\end{document}